# Resonant tunneling magnetoresistance in epitaxial metal-semiconductor heterostructures

J. Varalda[1], A. J. A. de Oliveira[1], D. H. Mosca[2], J.-M. George[3], M. Eddrief[4], M. Marangolo[4], V. H. Etgens[4]

[1]*Departamento de Física – UFSCar, C. P. 676, 13565-905 São Carlos SP, Brazil*
[2]*Departamento de Física – UFPR, C. P. 19091, 81531-990 Curitiba PR, Brazil*
[3]*Unite Mixte de Physique CNRS/Thales Domaine Corbeville, 91404 Orsay Cedex, France*
[4]*INSP,Institut des NanoSciences de Paris, UMR CNRS 7588, Universités Paris 6 et Paris 7 Campus Boucicaut – 140 rue de Lourmel – 75015 Paris, France*

We report on resonant tunneling magnetoresistance via localized states through a ZnSe semiconducting barrier which can reverse the sign of the effective spin polarization of tunneling electrons. Experiments performed on Fe/ZnSe/Fe planar junctions have shown that positive, negative or even its sign-reversible magnetoresistance can be obtained, depending on the bias voltage, the energy of localized states in the ZnSe barrier and spatial symmetry. The averaging of conduction over all localized states in a junction under resonant condition is strongly detrimental to the magnetoresistance.

Magnetic tunnel junctions (MTJs) consisting of two ferromagnetic electrodes separated by a thin tunnel barrier are one realistic way to build new and useful nanoelectronics devices such as magnetic random access memories, ultrahigh density magnetic read heads, picotesla field sensors as well as quantum computing units [1-5]. The tunneling magnetoresistance (TMR) is the change in tunneling current associated with the relative alignment of electrode magnetizations, written as $(G_P–G_{AP})/(G_{AP}+G_P) = P_1P_2$ [6]. Here, $G_P$ and $G_{AP}$ are the conductance for the parallel and antiparallel alignment whereas $P_1$ and $P_2$ are the spin polarization of the density of states at the Fermi energy of the two ferromagnets (FM).

In recent years, MTJs with semiconductor barriers have been widely investigated [1-6], but significant TMR values are observed only at relatively low temperatures, except for the recent results on MTJs with ZnSe barriers by Jiang et al. [7], where a TMR of about 10% (using the standard definition $(G_P–G_{AP})/G_{AP}$) has been reported at room temperature. MTJs with iron FM layer and ZnSe barriers are the strongest candidates for TMR studies and applications due to several advantageous conditions: favorable matching of Fe and ZnSe lattices [8], stable chemistry and magnetism at the interface [9], coherent spin lifetime as long as a fraction of a microsecond in n-type undoped ZnSe [10], electronic pinning of the Fe–Fermi level position at 1.6 eV above the valence–band maximum with a corresponding Schottky-barrier height of 1.1 eV [11], and a carrier concentration dominated by electrons [12] in a well-known electronic band structure including impurity/defect states in the energy band gap $E_G = 2.7$ eV [13,14]. Furthermore, several theoretical investigations have recently predicted that the conductance of Fe/ZnSe/Fe MTJs, with reasonably thick ZnSe barriers, could exhibit TMR as large as 100% due to spin-dependent tunneling into a matching electronic band [15]. However, the interactions of tunneling electrons with the electronic structure of the barrier, such as interfacial and mid-gap states, have not been taken into account. Bratkovsky [16] demonstrated theoretically that impurity-assisted tunneling decreases the TMR. More recently, Tsymbal et al. [5,6] described similar detrimental effects caused by mid-gap localized states in a series of Ni/NiO/Co nanojunctions.

In this letter, we demonstrate experimentally that an averaged resonant tunneling mechanism via mid-gap localized states generated by small levels of disorder inside the barrier reduces dramatically the TMR in Fe/ZnSe/Fe MTJs despite their high structural quality. When the tunneling electron energy driven by the bias voltage matches the energy of localized states (resonance), an inversion of the effective TMR spin polarization can occur. The phenomenon depends on the spatial symmetry of the electronic spin-dependent leak rates from the electrodes. Even if we don't report about room temperature TMR we will show, in this letter, how temperature can play an important role in defect-assisted resonant tunneling magnetoresistance. These findings bring a more general understanding of conditions favoring spin injection into an n-type semiconductor. It turns out that the averaged resonant tunneling via defect states generated by disorder in the barrier close to the chemical potential (~kT) appears to be a very general problem and has to be taken into account for hybrid FM/SC devices.

High quality epitaxial Fe (14nm)/ZnSe(8nm)/Fe(6nm) heterostructures were grown on a thin pseudomorphic ZnSe epilayer deposited on a GaAs buffer layer grown on a GaAs(001) substrate in a multichamber molecular beam epitaxy (MBE) system [9,11]. The best growth quality was obtained for Fe layers grown on a c(2x2) Zn- rich surface at 200 °C at a rate of ~0.14 nm/s (base pressure < 3 x $10^{-10}$ mbar). Details of the Fe growth on ZnSe(001) and of thin film properties are given elsewhere [9,11]. The Fe and ZnSe display chemically homogeneous (pinhole-free) epilayers with smooth surfaces and interfaces from atomic to larger scale. The semiconducting ZnSe barriers show, however, a small thickness fluctuation and some unavoidable growth defects (stacking faults and antiphase boundaries). This disorder may result in a broadening of conduction and valence bands and can also create localized states inside the ZnSe band gap, leading to conduction through tunneling-assisted mechanisms [5].

Next, 144 MTJs were fabricated simultaneously on a 1 cm$^2$ piece of sample by photolithography with cross sections from 24 to 380 μm$^2$ for transport measurements with current-perpendicular-to-the-plane geometry. The DC current–voltage characteristics of a MTJ between the bottom and top Fe layers were measured using a voltage source four-probe method in a continuous He flow cryostat between 4 and 300 K with an in-plane magnetic field ranging up to 6 kOe. Magnetization measurements were also performed on pieces of the unpatterned samples using a SQUID magnetometer, with applied magnetic field parallel to the sample surface.

Only a dozen MTJs in the ensemble, with areas varying from 24 μm$^2$ to 64 μm$^2$, have displayed a measurable TMR effect. We will show the results for three of them that represent well the observed phenomenon. They are: MTJ-A, a 24 μm$^2$ junction that displayed a positive TMR variation; MTJ-B, a 64 μm$^2$ junction that displayed negative TMR variation and; MTJ-C, another 24 μm$^2$ junction from the same ensemble that displayed both positive and negative TMR variation depending on the applied bias.

Figure 1(a) displays a typical magnetization hysteresis loop for the Fe/ZnSe/Fe heterostructure. The saturation magnetization of Fe (1710 G) is reached at large fields while a clear plateau of magnetization is observed between coercive fields of the top and bottom Fe layers. Figure 1(b) shows the resistance-area product as a function of the applied magnetic field at 10 K for the MTJ-A (24 μm$^2$). The relative change of junction resistance from parallel to antiparallel alignment of the magnetization coincides well with the switching fields of the magnetic moments of the two electrodes giving rise to a positive TMR that rapidly saturates at high field [cf. inset of Figure 1(b)]. We observe an important evolution of the resistance-area product RS that is around one order of magnitude higher for the MTJ-B (64 μm$^2$) (Figure1(c)). In both cases the RS value is much smaller than those typically found for insulating barriers [17]. A negative TMR is observed superimposed on a negative high-field MR contribution that is not related with a quadratic negative MR observed below 10 kOe in n-type ZnSe [14]. The hopping-assisted conduction along localized states under the drag of magnetic force was identified as a possible explanation for this high-field negative MR [18]. It may then result that the high-field negative MR and low-field negative TMR are distinct phenomena related to the localized states in the barrier.

The TMR measurement for the MTJ-C (24 μm$^2$) junction at 30 K is shown in Figure 2(a) and (b) at bias voltages of U = +0.5 V and U = +1.1 V, respectively. Positive and negative TMR are clearly visible depending on applied bias voltage meaning that the spin polarized electrons are injected into ZnSe from one electrode and are detected by the other electrode with normal and inverted spin polarization.

Despite the different magnitudes, the resistances of MTJ-A, MTJ-B and MTJ-C [cf. Figure 3(a)] exhibit an increase with decreasing temperature. This characteristic is essential to confirm the pinhole-free behavior as demonstrated by Akerman et al. [19]. The current versus voltage I(V) curves of these junctions exhibit a non-linear behavior below 30 K, as shown in Figure 3(b). Non-parabolic dI/dV curves are observed (inset Fig. 3(b)), indicating that the conductance is not dominated by direct tunneling processes at low temperature.
The TMR sign reversal was first observed by de Teresa et al. [20] on LSMO/SrTiO$_3$/Co where a strong asymmetry of the density of states of the LSMO electrode, associated with SrTiO$_3$/Co interface hybridization lead to negative TMR. In a recent report, Tsymbal et al. [6] observed the TMR sign reversal for Ni/NiO/Co nanojunctions and explained it in terms of resonant tunneling via mid-gap localized states in the barrier. Incidentally, the phenomenon observed by de Teresa et al. [20] seems unlikely in our sample since (i) the electrodes are both Fe and (ii) in some samples a positive TMR is observed (Figure 4a). A straightforward analogy can be made between our TMR versus bias results (Figure 4) and those of Tsymbal et al. [6], where resonant tunneling via localized states in the barrier close to the Fermi level also leads to an inversion of the TMR. Remembering that coherent direct tunneling connecting the two electrodes is unlikely according to our transport measurements, we start our analysis considering solely the resonance tunneling via localized states and the exponential dependence on the spatial position of defect states within the barrier.

According to the Landauer-Büttiker formula [21], the tunnel conductance is proportional to the transmission coefficient. Considering the tunnel conductance per spin channel $G_d$ as a function of energy E in the Breit-Wigner form given by [6]

$$G_d(E) = \frac{4e^2}{h} \frac{\Gamma_1 \Gamma_2}{(E-E_d)^2 + (\Gamma_1+\Gamma_2)^2} \quad (1)$$

where $E_d$ is the energy of the defect state, and $\Gamma_1/h$ and $\Gamma_2/h$ are the leak rates of an electron from the defect state to the bottom and top Fe electrodes, which are assumed for simplicity to be [6] $\Gamma_1 \propto \rho_1 \exp[-2\kappa x]$ and $\Gamma_2 \propto \rho_2 \exp[-2\kappa(d-x)]$ where $\rho_1$ and $\rho_2$ are the densities of states of the electrodes 1 (bottom) and 2 (top) and $d$ is the position of defect within the barrier. Off resonance, when $|E - E_d| \gg (\Gamma_1 + \Gamma_2)$, the latter assumption leads to $G_d \propto \rho_1\rho_2$ and TMR = $P^2$ as in Jullière's model with P being the polarization of the Fe electrodes. At resonance, when $E = E_d$, there are two possibilities: 1) the defect state is at the center of the barrier and $\Gamma_1 = \Gamma_2$; 2) the position of the defect is asymmetric and $\Gamma_1 \neq \Gamma_2$. In the first case, the conduction is maximum (transmittance is maximal), $G_d$ is still proportional to $\rho_1\rho_2$ and the TMR is positive and equals $P^2$. In the second case, if $\Gamma_1 \gg \Gamma_2$ (defect near electrode 1) or if $\Gamma_1 \ll \Gamma_2$ (defect near electrode 2), the conductance (1) is inversely proportional to the density of states of electrodes ($G_d \propto \rho_2/\rho_1$ or $G_d \propto \rho_1/\rho_2$) and one finds TMR = $-P^2$ [6]. For leak rates $\Gamma_1 \neq \Gamma_2$, the conductance still has a maximum at $E_d$ but the transmittance is lower when compared with the case $\Gamma_1 = \Gamma_2$.

The normalized TMR as a function of bias voltage for MTJ-A and MTJ-B is shown in Figures 4(a) and (b), respectively. The solid line in Figure 4(a) shows the calculated positive TMR from Eq. (1) with $E = eU + E_F$. Experimental data are well reproduced by considering an ideal Fe polarization of 44% [5] at each interface and $E_F = 1.6$ eV, as measured for the crystalline Fe/ZnSe contact [11], with $E_d = 45$ meV above $E_F$ and $\Gamma = \Gamma_1 = \Gamma_2 = 110$ meV. The defect states in ZnSe are distributed over the entire band gap region; however, the defect state density is considerably higher around the energy of 1.6 eV above the valence–band maximum [13]. In this case, the defect state has energy of 45 meV above $E_F$. The similar leak rates corroborate the assumption that defect states stand in the middle of the barrier. The variation of the TMR versus bias voltage for MTJ-B, which displays inverse TMR is controlled by the (spatial) position and width of the resonant energy state. The solid line in Figure 4(b) represents the best fit of (1) with $E = eU + E_F$ using the following parameters: $E_d = E_F$, $\Gamma_1 = 93$ meV and $\Gamma_2 = 26$ meV. In this case, the defect state energy coincides with $E_F$. A more abrupt change of the negative TMR versus bias voltage results from the smaller resonant widths. The resonance widths for MTJ-A and MTJ-B are about 220 meV and 120 meV respectively. These values are in good agreement with the value of ~180 meV reported for disordered ZnSe [13]. From the leak rates, $\Gamma_1 = 93$ meV and $\Gamma_2 = 26$ meV, one can estimate an average transfer time of electrons in the impurity states of about $\hbar/\Gamma \sim 10^{-14}$ sec.

It is interesting to note that the resistance values are strongly dependent on the spatial symmetry of defects. Negative TMR comes from defect states more asymmetric with respect to the electrodes, thereby resulting in $\Gamma_1 \neq \Gamma_2$ and higher RS values than in the case of positive TMR ($\Gamma_1 = \Gamma_2$).

The reversal of the TMR sign as function of bias (MTJ-C in Fig. 2) is related with the opening of new resonant conduction channels. For electrons in electrode 1 with $E = eU + E_F$, the optimal conduction channels are those for which the transmittance is maximum, in other words E ~ $E_d$ (resonance). If the electrode energy is increased by an amount $e\delta U$ larger than the energy width of the defect, the states with energy $E_d$ come out of resonance. Their transmittance decreases and these channels represent a minor contribution to the total conduction. Other channels with $E_d^* = E_d + e\delta U$ have their transmittance increased and become the main contribution to the total conduction. Thus, the TMR sign depends on the spatial symmetry of the defect states in the barrier of the new resonant channels and will be positive if $\Gamma_1 = \Gamma_2$ (symmetric position) or negative if $\Gamma_1 \neq \Gamma_2$ (asymmetric position).

Tsymbal et al. [6] have identified the nanometric areas of MTJs as an important experimental condition for avoiding impurity/defect-driven transport by a large number of local disorder configurations which determine an averaged reduction in the TMR. Surprisingly we have observed the resonant tunneling phenomena in planar junctions with cross section as large as tens of $\mu m^2$. This demonstrates the high quality (relatively low density of defects) of our Fe/ZnSe hybrid structures.

It is possible to simulate disorder in a real MTJ by averaging the conductance over the energies and positions of defects in the barrier. A rough estimate is obtained after integrating Eq. (1), assuming a homogeneous distribution of defects with a uniform density at $E_F$ and ZnSe barriers that are not too thin (i.e., a decay constant similar to the barrier thickness). Bratkovsky [16] demonstrated that TMR decreases to 4% even for very low defect levels (as low as $10^{-7}$ Å$^{-3}$eV$^{-1}$) in an insulator barrier with Fe electrodes. The reversal of the magnetoresistance still occurs at resonant conditions.

In this sense, the fact that the localized states are distributed over the entire band gap of crystalline semiconductor barriers with major density near the Fermi energy by a few meV could explain the difficulty in observing the TMR effect at room temperature in FM/SC hybrid structures. For each bias energy, the difference (E - $E_d$) ~ kT which means that as the temperature increases, more channels will be able to conduct and the current spin polarization disappears because of conduction averaging. The strong positive variation of the resistance with temperature decrease as observed for MTJ-C in Figure 3(a)) is a clear example of this situation. An increase in the thermal energy causes many other conduction channels to become active, which is detrimental to the TMR.

As a final comment, we remember that the conductance in the disordered junctions with large cross section is an average over a large number of channels. Each channel is dominated by disorder centers that correspond to defects with different energies and positions. This simply results in the suppression of the TMR as observed for MTJs with area larger than 64 $\mu m^2$. Thermal energy gives the same effect by opening new conduction channels. The small TMR previously reported [18] in epitaxial Fe/ZnSe/Fe MTJs can be understood in the light of the above explanations.

In conclusion, we have observed resonant tunneling via defect states in the barrier in MTJs with relatively large cross sections and small resistance-area products. These findings attest that in spite of the high quality of the hybrid structures a few defect states lead to resonant tunneling and can even invert the observed TMR. This can explain the reduction or extinction of the effective TMR We believe that a significant room temperature TMR effect can be obtained only if mid-gap defect states are removed. In addition, this study brings a more general understanding of the conditions for spin injection into an n-type semiconductor when defect resonant states in the barriers with energies close to the chemical potential are involved.

The authors acknowledge the bilateral CAPES-COFECUB program, FAPESP (grant 03/02804-8), CNPq and French Program Action Concertée Nanosciences-Nanotechologies for financial support.

**References**

[1] M. Jullière, Phys. Lett. **54A**, 225 (1975)

[2] T. Miyazaki and N. J. Tezuka, J. Magn. Magn. Mater. **139**, L231 (1995)

[3] J. S. Moodera et al., Phys. Rev. Lett. **74**, 3273 (1995)

[4] S. A. Wolf et al., Science **294**, 1488 (2001)

[5] E. Y. Tsymbal et al., J. Phys.: Condens. Matter. **15**, R109 (2003)

[6] E. Y. Tsymbal et al., Phys. Rev. Lett. **90**, 186602 (2003)

[7] X. Jiang, A. F. Panchula, and S. S. P. Parkin, Appl. Phys. Lett. **83**, 5244 (2003)

[8] G. A. Prinz et al., Appl. Phys. Lett. **48**, 1756 (1986) ; B. T. Jonker et al., J. Cryst. Growth **81**, 524 (1897)

[9] M. Marangolo et al. Phys. Rev. Lett. **88**, 217202 (2002)

[10] I. Malajovich et al., Phys. Rev. Lett. **84**, 1015 (2000)

[11] M. Eddrief et al., Appl. Phys. Lett. **81**, 4553 (2002)

[12] The n-type undoped ZnSe is mostly formed as a consequence of the compensation mechanisms by intrinsic defects.

[13] P. K. Lim, D. E. Brodie, Can. J. Phys. **55**, 1641 (1977)

[14] M. Vaziri, Appl. Phys. Lett. **65**, 2568 (1994)

[15] J. M. MacLaren et al., Phys. Rev. B **59**, 5470 (1999)

[16] A. M. Bratkovsky, Phys. Rev. B **56**, 2344 (1997). The possibility of resonant tunneling through barrier levels that have a spin-dependent density (see R. Jansen and J. C. Lodder, Phys. Rev. B 61 (2000) 5860) is not considered for defect levels.

[17] Gregg J. F., Petej I., Jouguelet E. and Dennis C., *Spin electronics – a review*, J. Phys. D: Appl. Phys. **35**, R121 (2002)

[18] D. H. Mosca, J. M. George, J. L. Maurice, A. Fert, M. Eddrief, V. H. Etgens, J. Magn. Magn. Mater. **226-230**, 917 (2001)

[19] J. J. Akerman, R. Escudero, C. Leighton, S. Kim, D.A. Rabson, R. W. Dave, J.M. Slaughter, I. K. Schuller, J. Magn. Magn. Mat. **240**, 86 (2002)

[20] J. M. de Teresa et al, Science **286**, 507 (1999)

[21] R. Landauer, IBM J. Res. Dev. **32**, 306 (1988); M. Büttiker, IBM J. Res. Dev. **32**, 317 (1988)

**Figure Captions**

**Figure 1** – (a) Magnetization of an Fe(14nm)/ZnSe(8nm)/Fe(6nm) heterostructure measured by SQUID at 10 K. Magnetic field dependence of the resistance–area product measured at 10 K for a junction with (b) 24 µm$^2$ (MTJ-A) measured at U = 70 mV and (c) 64 µm$^2$ (MTJ-B) measured at U = 500 mV. The inserts in (b) and (c) show the complete curves.

**Figure 2** – Bias inversion of the TMR measured at 30 K in the other junction with 24 µm$^2$ (MTJ-C).

**Figure 3** – (a) Temperature dependence of the resistance for MTJ-A, MTJ-B and MTJ-C measured under a bias voltage of 100 mV. (b) Current versus voltage curves of the junctions measured at 30 K. The insert shows the differential conduction.

**Figure 4** – Normalized magnetoresistance as a function of bias voltage for: (a) MTJ-A (TMR = +0.05) and (b) MTJ-B (TMR = –0.01 at low field).

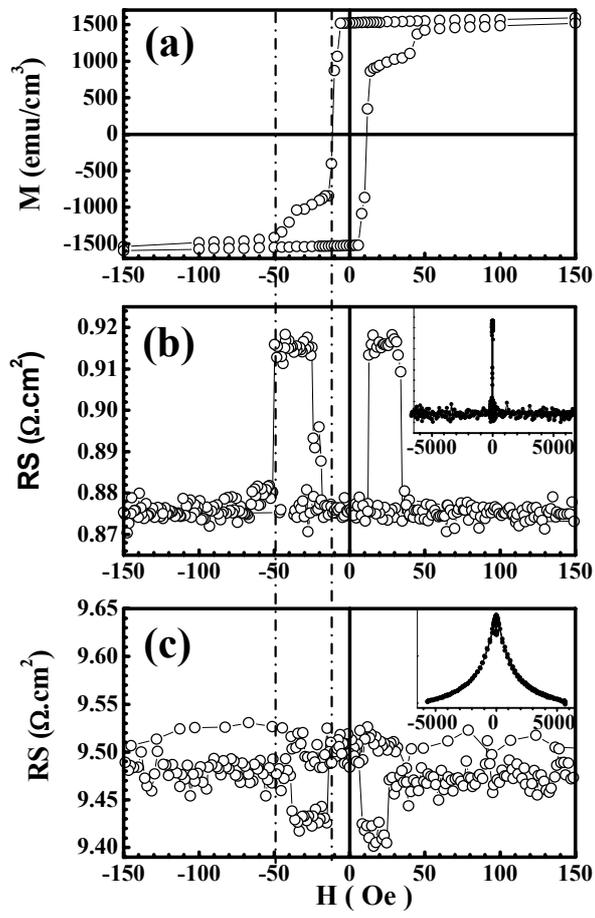

Varalda et al., Figure 1

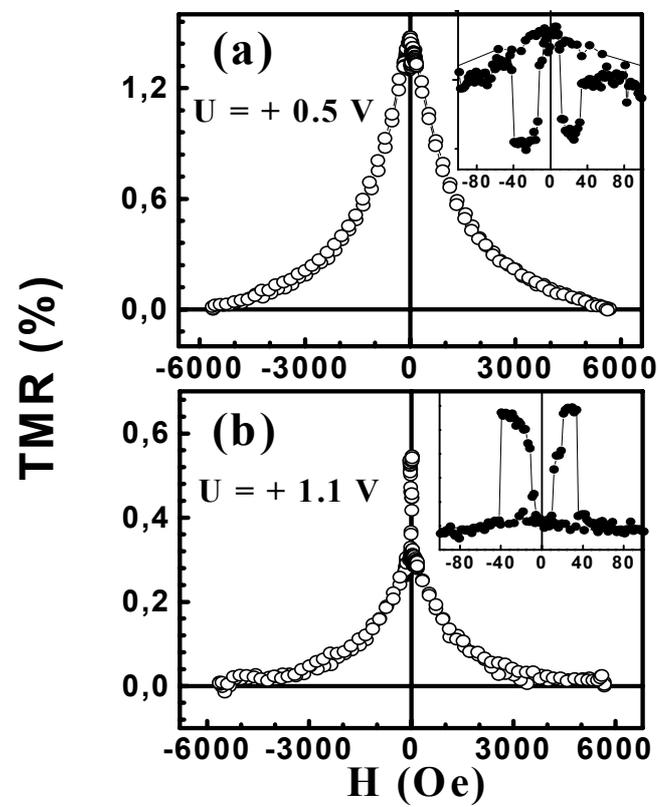

Varalda et al., Figure 2

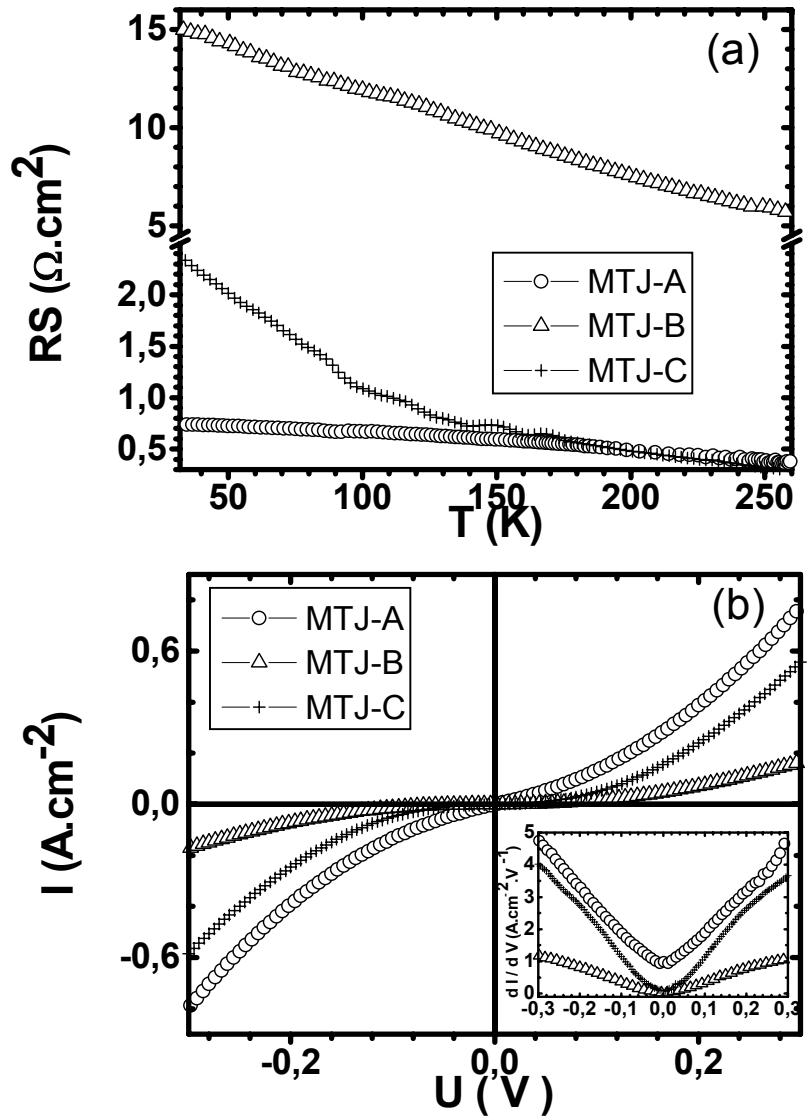

Varalda et al., Figura 3

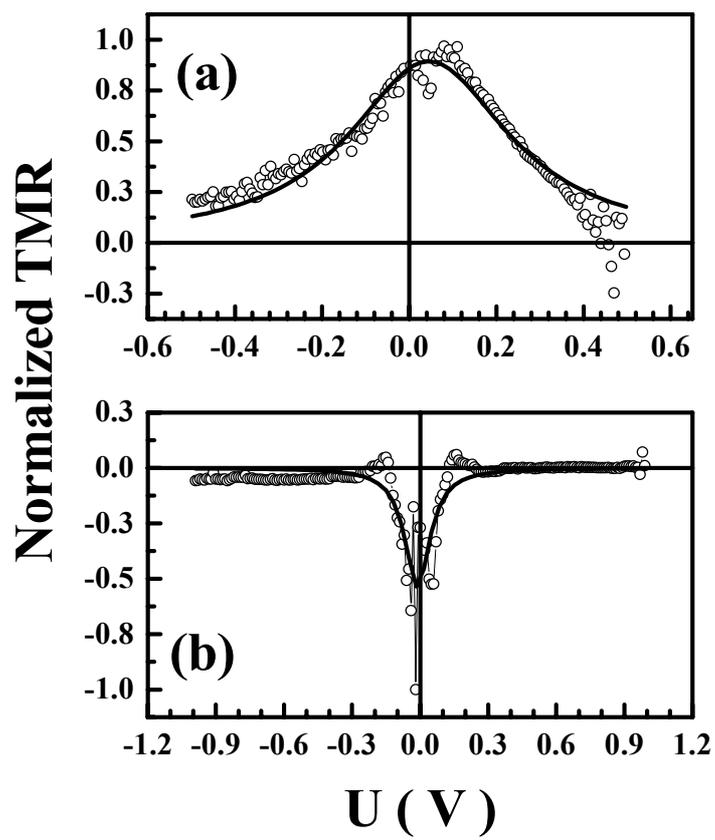

Varalda et al., Figure 4